\documentclass[reprint,superscriptaddress,nofootinbib,amsmath,amssymb,aps]{revtex4-2}
\usepackage{dcolumn,bm,graphicx,booktabs,float,url,hyperref,cleveref,physics,braket,qcircuit,tikz}
\usetikzlibrary{positioning}
\crefname{figure}{Fig.}{Figs.}
\crefname{equation}{Eq.}{Eqs.}
\crefname{section}{Sec.}{Sec.}
\Crefname{figure}{Figure}{Figures}
\Crefname{equation}{Equation}{Equations}
\Crefname{section}{Section}{Sections}

\DeclareMathOperator*{\argmax}{arg\,max}
\begin{document}

\preprint{APS/123-QED}
\title{Quantum-gate decomposer}

\author{Ken M. Nakanishi}
    \email{ken-nakanishi@g.ecc.u-tokyo.ac.jp}
    \affiliation{
        Institute for Physics of Intelligence,
        The University of Tokyo,
        Tokyo 113-0033, Japan.
    }
\author{Takahiko Satoh}
    \email{satoh@sfc.wide.ad.jp}
    \affiliation{
        Keio University Quantum Computing Center,
        Yokohama, Kanagawa 223-8522 Japan.
    }
    \affiliation{
        Graduate School of Science and Technology,
        Keio University,
        Yokohama, Kanagawa 223-8522 Japan.
    }
\author{Synge Todo}
    \email{wistaria@phys.s.u-tokyo.ac.jp}
    \affiliation{
        Department of Physics,
        The University of Tokyo,
        Tokyo 113-0033, Japan.
    }
    \affiliation{
        Institute for Physics of Intelligence,
        The University of Tokyo,
        Tokyo 113-0033, Japan.
    }
    \affiliation{
        Institute for Solid State Physics,
        The University of Tokyo,
        Kashiwa, 277-8581, Japan.
    }

\date{\today}

\begin{abstract}
    Efficient decompositions of multi-qubit gates are essential in NISQ applications, where the number of gates or the circuit depth is limited. This paper presents efficient decompositions of CCZ and CCCZ gates, typical multi-qubit gates, under several qubit connectivities. We can construct the CCZ gate with only four CZ-depth when the qubit is square-shaped, including one auxiliary qubit. In T-shaped qubit connectivity, which has no closed loop, we can decompose the CCCZ gate with 17 CZ gates. While previous studies have shown a CCCZ gate decomposition with 14 CZ gates for the fully connected case, we found only four connections are sufficient for 14 CZ gates' implementation. The search for constraint-sufficient decompositions is aided by an optimization method we devised to bring the parameterized quantum circuit closer to the target quantum gate. We can apply this scheme to decompose any quantum gates, not only CCZ and CCCZ. Such decompositions of multi-qubit gates, together with the newly found CCZ and CCCZ decompositions, shorten the execution time of quantum circuits and improve the accuracy of complex quantum algorithms on near future QPUs.
\end{abstract}

\maketitle

\section{Introduction}\label{sec:introduction}
    Efficient execution of complex qubit operations on a quantum processing unit (QPU) is a significant challenge in quantum computation, especially in today's NISQ devices~\cite{Preskill2018}, where the total error correction is not available. When executing a given quantum circuit, the quantum compiler replaces all \textit{high-level} multi-qubit quantum gates with sequences of \textit{primitive} qubit operations that can be directly executed on the QPU. For example, recent trapped-ion QPUs have high connectivity so that we can perform two-qubit gates on arbitrary pairs of qubits~\cite{wright2019benchmarking, yirka2020honeywell}. On the other hand, superconducting QPUs generally have sparse connectivity so that we can perform two-qubit gates only between physically connected qubits~\cite{kelly2018preview, vu2017ibm, caldwell2018parametrically}. In the following, we will call such one-, two-qubit operations ``primitive gates''.
    
    The set of primitive gates supported by QPUs are usually one- and two-qubit gates. In principle, arbitrary quantum gates can be built as sequences of one- and two-qubit gates~\cite{Deutsch1995-ax,Lloyd1995-fq,Bremner2002-eh}. For example, Barenco et al.\ and Cleve et al.\ have developed methods for building multi-qubit controlled unitary operations~\cite{Barenco1995-ro, Cleve1998-mr}. In the following, the conversion of quantum gates into a sequence of primitive gates is referred to as the ``decomposition''.

    \begin{figure}[t]
        \centering
        \includegraphics[width=\linewidth]{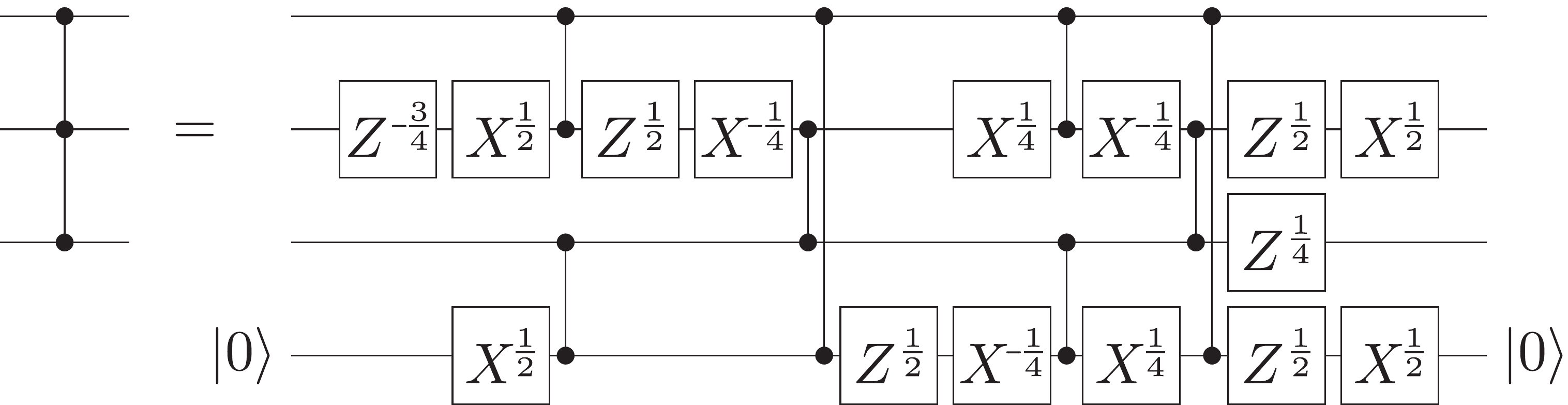}
        \caption{
            CCZ decomposition into CZ and one-qubit gates in the case of square-shaped qubit connectivity (third row in \Cref{tb:ccz}). $X^a := R_X\qty(a\pi)$, $Z^a := R_Z\qty(a\pi)$. The CCZ gate can be decomposed with only four CZ-depth.
            The qubit at the bottom of the right side is used as an auxiliary qubit. 
        }
        \label{fig:ccz13}
    \end{figure}
    
    To execute quantum algorithms on QPU efficiently with high accuracy, we should consider the following four points~\cite{khatri2019quantum,amy2013meet}:
    \begin{itemize}
    \item\textbf{Types of two-qubit primitive gates:}
    
        A most typical two-qubit gate in a quantum algorithm is the CNOT gate, but not all QPUs can execute this gate directly. Available two-qubit primitive gates, such as the iSWAP gate~\cite{PhysRevA.102.022619} and the CZ gate~\cite{PhysRevA.87.022309}, vary depending on the QPU's Hamiltonian.

    \item\textbf{Qubit connectivity:}
    
        In many architectures, QPUs can perform two-qubit operations directly only between limited qubit pairs. The topology or the graph, which defines the set of qubit pairs that can interact directly, is called ``qubit connectivity''.

    \item\textbf{Execution time:}
    
        Execution time of quantum operations is determined by the ``depth'' of the circuit. To reduce the execution time of a quantum algorithm, we have to decompose quantum operations into as short a sequence of primitive gates as possible by considering that qubit operations on non-overlapping sets of qubits can be executed simultaneously. Especially, the depth of two-qubit primitive gates is an essential metric for designing efficient decomposition.

    \item\textbf{Total number of gates:}
    
        In QPUs, the amount of noise is determined by the total number, or the ``count'', of the primitive gates, especially that of two-qubit primitive gates, as two-qubit primitive gates are often much noisier than one-qubit gates. To reduce the noise, therefore, the two-qubit count after the decomposition should be as small as possible.
    \end{itemize}
    
    To find efficient decompositions of given quantum gates, we have devised an optimization method for parameterized quantum circuits. There are many previous studies on the methods for finding efficient decomposition of quantum gates~\cite{khatri2019quantum,venturelli2018compiling,cincio2018learning,zahedinejad2016designing,booth2018comparing,maslov2016advantages,martinez2016compiling,younis2020qfast,amy2013meet}. In the present paper, we propose a method to search for a quantum gate decomposition using a classical computer, taking moderate-size qubit gates that fit the current classical computers as the targets. Several methods have been devised to find the quantum gate decomposition by optimizing the rotation parameters in a parameterized quantum circuit. The steepest descent method, BFGS method, and simulated annealing method have been used to optimize the parameters~\cite{cincio2018learning,martinez2016compiling,khatri2019quantum}. We have devised a technique that significantly reduces the computational complexity of this parameter optimization based on the method proposed in~\cite{Nakanishi2020}. Using this method, we can rapidly optimize the circuit parameters for a given parameterized quantum circuit to match the target quantum gate if possible. By examining different parameterized quantum circuits, we can find an efficient decomposition of the target quantum gate. It is also possible to replace the optimization part of the existing studies with our algorithm. Moreover, it can help us to check whether the already known quantum gate decomposition is optimal or not.

    We apply our optimization method to find gate decompositions of CCZ and CCCZ under different qubit connectivities. The CCZ and CCCZ gates are multi-qubit gates commonly used in quantum algorithms. As shown in \cref{tb:cccz}, various efficient quantum-gate decompositions have been found for CCZ and CCCZ under some connectivities~\cite{Schuch2003-ze, Schuch_undated-yw}. We found more quantum-gate decompositions on different connectivities. We assume only the CZ gate as a two-qubit primitive gate. We find that when the qubit connectivity is square-shaped and contains one auxiliary qubit, the CCZ gate can be decomposed with CZ-depth (defined in \cref{sec:findings}) only four. The same is true for the Toffoli gate, as it can be made with a CCZ gate and two one-qubit gates. In the case of T-shaped qubit connectivity, which has no closed loop, the CCCZ gate can be decomposed using 17 CZ gates. The smallest number of CZ gates currently known to build a CCCZ gate is 14, which has been realized for the fully connected case. We, however, found that this lower bound can be achieved under only four connections between qubits. These newly found decompositions are expected to shorten the execution time of quantum circuits and improve the accuracy of quantum algorithms on NISQ devices.
    
    This paper is organized as follows: We present the efficient decompositions of CCZ and CCCZ gates we found~(\cref{sec:findings}). Next, we describe the overall framework for searching for quantum gate decompositions~(\cref{sec:experiment}). Then, we explain the optimization method we used for finding optimal rotation angles of a parameterized quantum circuit~(\cref{sec:method}).
    Finally, we summarize the present study and give an outlook of future works~(\cref{sec:conclusion}).
    
\section{Main results: decompositions of CCZ and CCCZ}\label{sec:findings}
\begin{figure*}[tbp]
        \begin{minipage}[h]{\linewidth}
            \vspace{5mm}
            \centering
            \includegraphics[width=\linewidth]{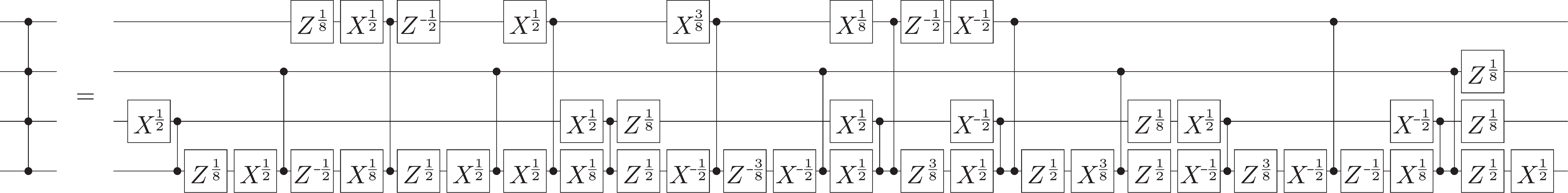}
            \caption{
                CCCZ gate decomposition into CZ and one-qubit gates in the case of T-shaped qubit connectivity (second row in \Cref{tb:cccz}). $X^a := R_X\qty(a\pi)$, $Z^a := R_Z\qty(a\pi)$. The CCCZ gate requires only 17 CZ gates in the case of T-shaped qubit connectivity.
            }
            \label{fig:cccz02}
        \end{minipage}

        \begin{minipage}[h]{\linewidth}
            \vspace{5mm}
            \centering
            \includegraphics[width=\linewidth]{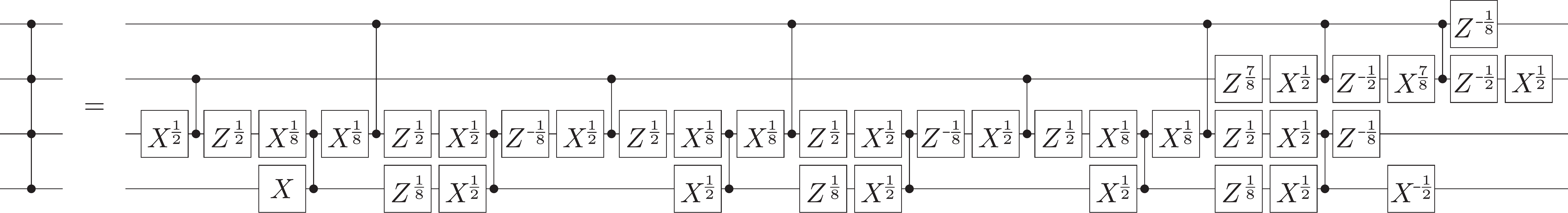}
            \caption{
                CCCZ gate decomposition into CZ and one-qubit gates. $X^a := R_X\qty(a\pi)$, $Z^a := R_Z\qty(a\pi)$. 
                The smallest number of CZ gates currently known to build a CCCZ gate is 14, which has been realized for the fully connected case.
                However, this lower bound can be achieved under only four connections between qubits (fourth row in \Cref{tb:cccz}).
            }
            \label{fig:cccz04}
        \end{minipage}
    \end{figure*}

    First, we present efficient decompositions of CCZ and CCCZ gates, typical three- and four-qubit gates, under different qubit connectivities. Here, we assume that the CZ gate is the only two-qubit primitive gate. We define the ``CZ-count'' as the total number of the CZ gates and the ``CZ-depth'' as the depth of the circuit, ignoring all one-qubit gates. Our main focus is on reducing the ``CZ-count'' and/or the ``CZ-depth'' of the decomposed circuit. \Cref{tb:ccz,tb:cccz} summarize the CCZ and CCCZ decompositions, respectively, for various connectivities. The quantum gate decompositions we found are shown in \cref{fig:ccz13,fig:cccz02,fig:cccz04}.

    \begin{table}[tbp]
        \centering
        \begin{tabular}{c|c|c|l}
            Connection & CZ-count & CZ-depth & Reference\\\hline\hline
            
            \rule{0pt}{8.5mm}
            \begin{tikzpicture}
                \tikzstyle{every node}=[draw,shape=circle,fill=black]
                \path(0,0)node(p0){}(.5,0)node(p1){}(.25,.4325)node(p2){};
                \draw(p0)--(p1)(p0)--(p2)(p1)--(p2);
            \end{tikzpicture}
             & 6 & 6 & Textbook implementation\\\hline
             
            \rule{0pt}{4.5mm}
            \begin{tikzpicture}
                \tikzstyle{every node}=[draw,shape=circle,fill=black]
                \path(0,0)node(p0){}(.5,0)node(p1){}(1,0)node(p2){};
                \draw(p0)--(p1)(p1)--(p2);
            \end{tikzpicture}
             & 8 & 8 & \citet{gwinner2020benchmarking}\\\hline
             
            \rule{0pt}{9mm}
            \begin{tikzpicture}
                \tikzstyle{every node}=[draw,shape=circle,fill=black]
                \path(0,0)node[fill=white](p0){}(.5,0)node(p1){}(.5,-.5)node(p2){}(0,-.5)node(p3){};
                \draw(p0)--(p1)(p1)--(p2)(p2)--(p3)(p3)--(p0);
            \end{tikzpicture}
             & 8 & $\bm{4}$ & \cref{fig:ccz13} \\\hline

        \end{tabular}
        \caption{
            CCZ gate decomposition into CZ and one-qubit gates. Bold numbers are new ones we found. In the Connection column, the black circles represent the target qubits, the white circles represent the auxiliary qubits, and the lines represent the qubit connectivities.
        }
        \label{tb:ccz}
    \end{table}

    \begin{table}[htbp]
        \centering
        \begin{tabular}{c|c|c|l}
            Connection & CZ-count & CZ-depth & Reference\\\hline\hline
            \rule{0pt}{9mm}
            \begin{tikzpicture}
                \tikzstyle{every node}=[draw,shape=circle,fill=black]
                \path(0,0)node(p0){}(0,.5)node(p1){}(.5,0)node(p2){}(.5,.5)node(p3){};
                \draw(p0)--(p1)(p0)--(p2)(p0)--(p3)(p1)--(p2)(p1)--(p3)(p2)--(p3);
            \end{tikzpicture}
             & 14 & 8 & \citet{Schuch_undated-yw} \\\hline
             
            \rule{0pt}{9mm}
            \begin{tikzpicture}
                \tikzstyle{every node}=[draw,shape=circle,fill=black]
                \path(0,0)node(p0){}(-.5,0)node(p1){}(0,-.5)node(p2){}(.5,0)node(p3){};
                \draw(p0)--(p1)(p0)--(p2)(p0)--(p3);
            \end{tikzpicture}
             & $\bm{17}$ & $\bm{17}$ & \cref{fig:cccz02} \\\hline

            \rule{0pt}{9mm}
            \begin{tikzpicture}
                \tikzstyle{every node}=[draw,shape=circle,fill=black]
                \path(0,0)node(p0){}(.5,0)node(p1){}(.5,-.5)node(p2){}(0,-.5)node(p3){};
                \draw(p0)--(p1)(p1)--(p2)(p2)--(p3)(p3)--(p0);
            \end{tikzpicture}
             & 16 & 8 & \citet{Schuch_undated-yw} \\\hline

            \rule{0pt}{9mm}
            \begin{tikzpicture}
                \tikzstyle{every node}=[draw,shape=circle,fill=black]
                \path(.5,0)node(p0){}(0,0)node(p1){}(-.4325,-.25)node(p2){}(-.4325,.25)node(p3){};
                \draw(p0)--(p1)(p1)--(p2)(p2)--(p3)(p3)--(p1);
            \end{tikzpicture}
             & $\bm{14}$ & - & \cref{fig:cccz04} \\\hline

            \rule{0pt}{4.5mm}
            \begin{tikzpicture}
                \tikzstyle{every node}=[draw,shape=circle,fill=black]
                \path(0,0)node(p0){}(.5,0)node(p1){}(1,0)node(p2){}(1.5,0)node(p3){};
                \draw(p0)--(p1)(p1)--(p2)(p2)--(p3);
            \end{tikzpicture}
             & 18 & 12 & \citet{Schuch_undated-yw} \\\hline
        \end{tabular}
        \caption{
            CCCZ gate decomposition into CZ and one-qubit gates. Bold numbers are new ones we found. In the Connection column, the black circles represent the target qubits of CCCZ, and the lines represent the qubit connectivities.
        }
        \label{tb:cccz}
    \end{table}
    
\section{Searching for quantum-gate decompositions}\label{sec:experiment}
    This section describes the overall framework for searching for quantum-gate decompositions. In the following, we denote the $X$ ($Z$) gate with rotation angle $\theta$ as $R_{X}(\theta)$ ($R_{Z}(\theta)$).

    \subsection{Parameterized quantum circuit generation}
    We start our search by determining the following:
    \begin{itemize}
        \item Type of two-qubit primitive gate to use.
        \item QPU qubit connectivity.
        \item Initial two-qubit count or two-qubit depth of the circuit.
    \end{itemize}
    First, we enumerate all possible sequences of two-qubit primitive gates according to these conditions. Then, we attach parameterized one-qubit gates, more specifically, $R_Z(\theta)$--$R_X(\theta')$--$R_Z(\theta'')$, before and after each two-qubit primitive gate. (Especially when we consider the CZ gate as two-qubit primitive gates, we can use $R_Z(\theta)$--$R_X(\theta')$ instead except at the end of the circuit. This is because CZ and $R_Z$ are commutative with each other, and thus omitting either one of the $R_Z$s before or after CZ does not spoil the representability of the parameterized circuit.) This way, we generate all the possible parameterized quantum circuits under the assumed conditions.

    \subsection{Exhaustive optimization of all prepared circuits}
    Next, we perform optimization of rotation angles in the parameterized quantum circuits we prepared. Details of the optimization will be presented in \cref{sec:method}. Since the optimization of rotation angles may stop at a local optimum, we ran the optimization several to several thousand times for each parameterized quantum circuit. The goal is achieved if the optimization finds a parameterized quantum circuit that matches the target quantum gate. Otherwise, we increase the two-qubit count or the two-qubit depth by one and start over again.
    
    \subsection{Further circuit simplification}
    Once a parameterized quantum circuit with some optimal rotation angles is found, one can further simplify the circuit by reducing the number of one-qubit gates in the primitive gate sequence. We perform thousands to millions of additional optimization runs starting from different random initial rotation angles for the parameterized quantum circuit that has converged to the target quantum gate. Then we examine the distribution of each rotation angle after convergence. We choose rotation gates whose rotation angle after convergence is often near zero or distributes almost evenly and remove one of such rotation gates by fixing its rotation angle to zero. This operation is repeated until there remain no more rotation gates that can be removed safely.

\section{Sequential optimization of rotating gates}\label{sec:method}
    In this section, we describe our sequential optimization algorithm for finding optimal rotation angles of a parameterized quantum circuit, which is partially based on the method proposed in~\cite{Nakanishi2020}. Our method is also applicable for a circuit with auxiliary qubits. We design this algorithm to run on a classical computer. In \cref{sec:objective_function}, we explain the property to be satisfied by the rotation gates in a parameterized quantum circuit we optimize and define the objective function \cref{eq:objective_function}. Then, in \cref{sec:flow}, we explain the flow of the sequential optimization method to maximize the objective function \cref{eq:objective_function}.

\subsection{Objective function}\label{sec:objective_function}

    We assume that all rotation gates $R_A(\theta)$ in a parameterized quantum circuit are  expressed as
    \begin{equation}\label{eq:rotation_gate}
        R_A(\theta)=\exp(-\frac{i\theta}{2}A)
    \end{equation}
    with $A$ satisfying the following condition:
    \begin{equation}\label{eq:rotation_gate_condition}
        A^2=I.
    \end{equation}
    We also assume that the input quantum states are in the space spanned by $D$ mutually orthogonal quantum states $\qty{\ket{\Phi_d}}_{d=1}^{D}$, and $P$ is defined as follows:
    \begin{equation}\label{eq:g_defP}
        P := \sum_{d=1}^D \ketbra{\Phi_d}{\Phi_d}.
    \end{equation}
    If the input space is the whole Hilbert space of the prepared $c$ qubits, $P$ is the $2^c$-dimensional identity matrix.
    
    Let $V_T$ be the target quantum gate we want to decompose, and $V$ be a parameterized quantum circuit. The goal of the present algorithm is to optimize the parameters in $V$ so that the output $\qty{V\ket{\Phi_d}}_{d=1}^{D}$ becomes closer to $\qty{V_T\ket{\Phi_d}}_{d=1}^{D}$, and eventually becomes identical except for the global phase common to all $D$ states. This optimization problem is identical to finding $V$ such that $f(V)=0$ by minimizing $f(V)$ defined by
    \begin{equation}\label{eq:def_f}
        f(V) := \min_\phi \norm{e^{i\phi}V_T^\dag VP - P}_F^2,
    \end{equation}
    where $\norm{\cdot}_F$ denotes the Frobenius norm.
    \Cref{eq:def_f} can be transformed into the following equation:
    \begin{equation}\label{eq:trU}
        f(V) = 2D - 2\qty|\tr[V_T^\dag VP]|.
    \end{equation}
    (See appendix~\ref{app:trU} for the detailed derivation.)
    Thus, minimizing $f(V)$ until $f(V)=0$ is equivalent to maximizing
    \begin{equation}\label{eq:objective_function}
        \qty|\tr[V_T^\dag VP]|^2
    \end{equation}
    until 
    \begin{align}
    \qty|\tr[V_T^\dag VP]|^2=D^2.
    \end{align}

\subsection{Rotation-angle optimization flow}\label{sec:flow}
    \begin{figure}[tbp]
        \centering
        \includegraphics[width=\linewidth]{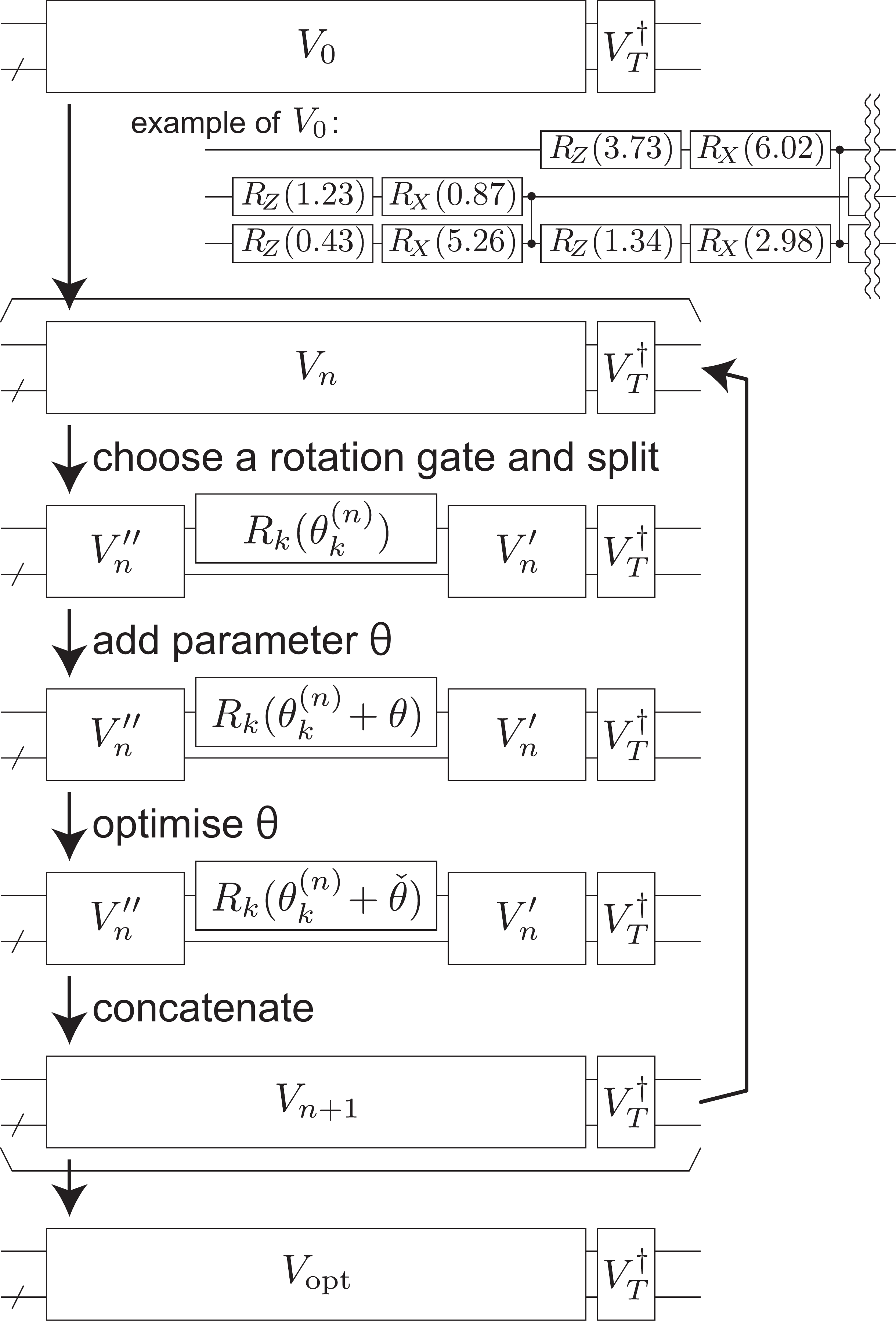}
        \caption{
            Rotation-angle optimization flow. $V_0$ denotes the initial state of the parameterized quantum circuit where the rotation angles are initialized randomly.
        }
        \label{fig:flow}
    \end{figure}
    Next we describe how to optimize the parameterized quantum circuit sequentially. The optimization flow is shown in \cref{fig:flow}.
    Suppose that the parameterized quantum circuit $V$ has $K$ rotation gates.
    Then $V$ can be written as
    \begin{align}
        V(\theta_1,\ldots,\theta_K) = W_K R_K(\theta_K) \cdots W_1 R_1(\theta_1) W_0,
    \end{align}
    where $\theta_k$ ($k=1,\ldots,K$) is the rotation angle of the $k$-th rotation gate and $W_k$ ($k=0,\ldots,K$) denotes a fixed multi-qubit unitary gate or an identity gate.
    Note that $W^{-1}_k = W_k$ ($k=0,\ldots,K$).

    Let $V_n=V(\theta_1^{(n)},\ldots,\theta_K^{(n)})$ be the parameterized quantum circuit after $n$ optimization steps. At the $(n+1)$-th step, first we choose one of the rotation gates, say $R_k(\theta_k^{(n)})$, in the parameterized quantum circuit $V_n$. We rewrite the parameterized circuit $V_n$ as
    \begin{equation}\label{eq:splitU}
        V_n = V'_n R_k(\theta_k^{(n)}) V''_n,
    \end{equation}
    where
    \begin{align}
        V'_n &= W_K R_K(\theta_K^{(n)}) \cdots R_{k+1}(\theta_{k+1}^{(n)}) W_{k}, \\
        V''_n &= W_{k-1} R_{k-1}(\theta_{k-1}^{(n)}) \cdots R_{1}(\theta_1^{(n)}) W_0.
    \end{align}
    Now we define $\tilde{V}_n(\theta)$ as
    \begin{align}
        \tilde{V}_n(\theta) &:= V'_n R_{k}(\theta_k^{(n)}+\theta) V''_n.
    \end{align}
    Then $\qty|\tr[V_T^\dag \tilde{V}_n(\theta)P]|^2$ can be transformed as
    \begin{align}
        \qty|\tr[V_T^\dag \tilde{V}_n(\theta)P]|^2
        =& \frac{\qty|t_0|^2 - \qty|t_\pi|^2}{2} \cos\theta 
        + \frac{t_0 t_\pi^* + t_0^* t_\pi}{2} \sin\theta \notag\\
        &+ \frac{\qty|t_0|^2 + \qty|t_\pi|^2}{2} \label{eq:sinU}
    \end{align}
    with
    \begin{align}
        t_0 &:= \qty|\tr[V_T^\dag \tilde{V}_n(0)P]|^2 = \qty|\tr[V_T^\dag V_n P]|^2, \label{eq:t0} \\
        t_\pi &:= \qty|\tr[V_T^\dag \tilde{V}_n(\pi)P]|^2, \label{eq:tpi}
    \end{align}
    where $t^*$ denotes the complex conjugate of $t$. (See appendix~\ref{app:sinU} for the derivation.)
    From \cref{eq:sinU}, we can find $\check\theta$ that maximizes $\qty|\tr[V_T^\dag \tilde{V}_n(\theta)P]|^2$ as
    \begin{align} \label{eq:argmax_theta}
        \check\theta &:= \argmax_\theta\qty(\qty|\tr[V_T^\dag \tilde{V}_n(\theta)P]|^2)\notag\\
        &= \begin{cases}
            \arctan(\frac{2 t_0 t^*_\pi}{\qty|t_0|^2 - \qty|t_\pi|^2}) & \qty|t_0|^2 - \qty|t_\pi|^2 > 0 \\
            \arctan(\frac{2 t_0 t^*_\pi}{\qty|t_0|^2 - \qty|t_\pi|^2}) + \pi & \qty|t_0|^2 - \qty|t_\pi|^2 < 0,
            \end{cases}
    \end{align}
    which defines $V_{n+1}$ for the next iteration:
    \begin{equation} \label{eq:defVp}
        V_{n+1} := \tilde{V}_n(\check\theta).
    \end{equation}

    From \cref{eq:sinU,eq:t0,eq:defVp}, the following inequality can be derived:
    \begin{equation}
        \qty|\tr[V_T^\dag V_{n+1}P]|^2 \geq \qty|\tr[V_T^\dag V_nP]|^2,
    \end{equation}
    that is, the objective function~(\ref{eq:objective_function}) increases monotonically as the optimization proceeds. By updating each rotation gate $R_k(\theta_k)$ ($k=1,\ldots,K$) sequentially according to the above procedure, we can optimize $V$ so that $\qty|\tr[V_T^\dag VP]|^2$ is maximized. If $\qty|\tr[V_T^\dag VP]|^2$ is maximized until $D^2$, the parameterized quantum circuit becomes identical to the target quantum gate, meaning that we successfully find a decomposition of $V_T$.

\subsection{Reduction of computation complexity}\label{sec:save_time}
    To reduce the computation complexity, we can exploit the cyclic property of the trace. In the objective function at the $(n+1)$ step, we can bring the chosen rotation gate to the leftmost as
    \begin{equation}
        \qty|\tr[V_T^\dag \tilde{V}_n(\theta)P]|^2 = \qty|\tr[R_k(\theta_k^{(n)}+\theta)M_n]|^2
    \end{equation}
    where
    \begin{align}
        M_n = V''_n P V_T^\dag V'_n.
    \end{align}
    If we choose $R_{k-1}(\theta_{k-1})$ or $R_{k+1}(\theta_{k+1})$ as the target rotation gate of the next step, $M_{n+1}$ can be easily calculated as
    \begin{align}
        M_{n+1} = R_{k-1}(-\theta_{k-1}^{(n)}) W_{k-1} M_n R_k(\check\theta) W_{k-1}
    \end{align}
    or
    \begin{align}
        M_{n+1} = W_k R_k(\check\theta) M_n W_k R_{k+1}(-\theta_{k+1}^{(n)}),
    \end{align}
    respectively, where we use the following properties: $R^{-1}_k(\theta_k) = R_k(-\theta_k)$ and $W^{-1}_k = W_k$.
    In the present calculation, we choose the rotation gates in the following order:
    \begin{align*}
        K &\rightarrow K-1 \rightarrow \cdots \rightarrow 2 \rightarrow 1 \rightarrow 2 \rightarrow \cdots \rightarrow K-1 \rightarrow K \\ &\rightarrow K-1 \rightarrow \cdots.
    \end{align*}
    Using this technique, we can reduce the computational cost significantly compared to calculating $V_n$ from scratch at each step.

\section{Conclusion}\label{sec:conclusion}
    In the present paper, we presented efficient decompositions of the CCZ and CCCZ gates under some qubit connectivities. The CCZ and CCCZ gates are multi-qubit gates commonly used in quantum circuits. Using these decompositions, we can achieve more efficient and less error-prone quantum circuits, especially for QPUs with sparse connectivity, such as the superconducting QPUs.

    Contributing to this finding is our sequential optimization algorithm of a parameterized quantum circuit. By using our method, we can optimize the rotation angles for a given parameterized quantum circuit to achieve the target quantum gate. The present optimization method can be used not only for CCZ and CCCZ but also for arbitrary qubit gates. Since the method does not depend on any particular primitive gate set, it can be modified and applied to various QPUs. Using this method to find a suitable quantum gate decomposition for particular QPUs will reduce the execution time of quantum circuits on NISQ devices and improve the accuracy of quantum algorithms. It is also possible to replace the parameter optimization part of the existing gate decomposition research with this method. We believe that the present optimization method will be an essential tool for error reduction in the NISQ era. Not only that, but the analysis of the quantum gate decomposition found in the present study may provide hints for designing larger-scale multi-qubit gates in the future.

\begin{acknowledgments}
    This calculation has been done using NVIDIA GPGPU at the Institute for Physics of Intelligence ($i\pi$), School of Science, the University of Tokyo.
    KMN has been supported by KAKENHI No.20J13955 and is supported by the Daikin Endowed Research Unit: ``Research on Physics of Intelligence'', School of Science, the University of Tokyo.
    TS is supported by MEXT Quantum Leap Flagship Program Grant Number JPMXS0118067285 and JPMXS0120319794.
    ST acknowledges support by the Endowed Project for Quantum Software Research and Education, The University of Tokyo (\url{https://qsw.phys.s.u-tokyo.ac.jp/}).
\end{acknowledgments}

\bibliography{main}

\begin{thebibliography}{26}%
\makeatletter
\providecommand \@ifxundefined [1]{%
 \@ifx{#1\undefined}
}%
\providecommand \@ifnum [1]{%
 \ifnum #1\expandafter \@firstoftwo
 \else \expandafter \@secondoftwo
 \fi
}%
\providecommand \@ifx [1]{%
 \ifx #1\expandafter \@firstoftwo
 \else \expandafter \@secondoftwo
 \fi
}%
\providecommand \natexlab [1]{#1}%
\providecommand \enquote  [1]{``#1''}%
\providecommand \bibnamefont  [1]{#1}%
\providecommand \bibfnamefont [1]{#1}%
\providecommand \citenamefont [1]{#1}%
\providecommand \href@noop [0]{\@secondoftwo}%
\providecommand \href [0]{\begingroup \@sanitize@url \@href}%
\providecommand \@href[1]{\@@startlink{#1}\@@href}%
\providecommand \@@href[1]{\endgroup#1\@@endlink}%
\providecommand \@sanitize@url [0]{\catcode `\\12\catcode `\$12\catcode
  `\&12\catcode `\#12\catcode `\^12\catcode `\_12\catcode `\%12\relax}%
\providecommand \@@startlink[1]{}%
\providecommand \@@endlink[0]{}%
\providecommand \url  [0]{\begingroup\@sanitize@url \@url }%
\providecommand \@url [1]{\endgroup\@href {#1}{\urlprefix }}%
\providecommand \urlprefix  [0]{URL }%
\providecommand \Eprint [0]{\href }%
\providecommand \doibase [0]{https://doi.org/}%
\providecommand \selectlanguage [0]{\@gobble}%
\providecommand \bibinfo  [0]{\@secondoftwo}%
\providecommand \bibfield  [0]{\@secondoftwo}%
\providecommand \translation [1]{[#1]}%
\providecommand \BibitemOpen [0]{}%
\providecommand \bibitemStop [0]{}%
\providecommand \bibitemNoStop [0]{.\EOS\space}%
\providecommand \EOS [0]{\spacefactor3000\relax}%
\providecommand \BibitemShut  [1]{\csname bibitem#1\endcsname}%
\let\auto@bib@innerbib\@empty
\bibitem [{\citenamefont {Preskill}(2018)}]{Preskill2018}%
  \BibitemOpen
  \bibfield  {author} {\bibinfo {author} {\bibfnamefont {J.}~\bibnamefont
  {Preskill}},\ }\href {https://doi.org/10.22331/q-2018-08-06-79} {\bibfield
  {journal} {\bibinfo  {journal} {Quantum}\ }\textbf {\bibinfo {volume} {2}},\
  \bibinfo {pages} {79} (\bibinfo {year} {2018})}\BibitemShut {NoStop}%
\bibitem [{\citenamefont {Wright}\ \emph {et~al.}(2019)\citenamefont {Wright},
  \citenamefont {Beck}, \citenamefont {Debnath}, \citenamefont {Amini},
  \citenamefont {Nam}, \citenamefont {Grzesiak}, \citenamefont {Chen},
  \citenamefont {Pisenti}, \citenamefont {Chmielewski}, \citenamefont {Collins}
  \emph {et~al.}}]{wright2019benchmarking}%
  \BibitemOpen
  \bibfield  {author} {\bibinfo {author} {\bibfnamefont {K.}~\bibnamefont
  {Wright}}, \bibinfo {author} {\bibfnamefont {K.}~\bibnamefont {Beck}},
  \bibinfo {author} {\bibfnamefont {S.}~\bibnamefont {Debnath}}, \bibinfo
  {author} {\bibfnamefont {J.}~\bibnamefont {Amini}}, \bibinfo {author}
  {\bibfnamefont {Y.}~\bibnamefont {Nam}}, \bibinfo {author} {\bibfnamefont
  {N.}~\bibnamefont {Grzesiak}}, \bibinfo {author} {\bibfnamefont {J.-S.}\
  \bibnamefont {Chen}}, \bibinfo {author} {\bibfnamefont {N.}~\bibnamefont
  {Pisenti}}, \bibinfo {author} {\bibfnamefont {M.}~\bibnamefont
  {Chmielewski}}, \bibinfo {author} {\bibfnamefont {C.}~\bibnamefont
  {Collins}}, \emph {et~al.},\ }\href@noop {} {\bibfield  {journal} {\bibinfo
  {journal} {Nature communications}\ }\textbf {\bibinfo {volume} {10}},\
  \bibinfo {pages} {1} (\bibinfo {year} {2019})}\BibitemShut {NoStop}%
\bibitem [{\citenamefont {Yirka}(2020)}]{yirka2020honeywell}%
  \BibitemOpen
  \bibfield  {author} {\bibinfo {author} {\bibfnamefont {B.}~\bibnamefont
  {Yirka}},\ }\href
  {https://phys.org/news/2020-06-honeywell-built-highest-performing-quantum.html}
  {\bibinfo {title} {Honeywell claims to have built the highest-performing
  quantum computer available}} (\bibinfo {year} {2020})\BibitemShut {NoStop}%
\bibitem [{\citenamefont {Kelly}(2018)}]{kelly2018preview}%
  \BibitemOpen
  \bibfield  {author} {\bibinfo {author} {\bibfnamefont {J.}~\bibnamefont
  {Kelly}},\ }\href@noop {} {\bibfield  {journal} {\bibinfo  {journal} {Google
  Research Blog}\ }\textbf {\bibinfo {volume} {5}} (\bibinfo {year}
  {2018})}\BibitemShut {NoStop}%
\bibitem [{\citenamefont {IBM}(2017)}]{vu2017ibm}%
  \BibitemOpen
  \bibfield  {author} {\bibinfo {author} {\bibnamefont {IBM}},\ }\href
  {https://www-03.ibm.com/press/us/en/pressrelease/53374.wss} {\bibinfo {title}
  {Ibm announces advances to ibm quantum systems \& ecosystem}} (\bibinfo
  {year} {2017})\BibitemShut {NoStop}%
\bibitem [{\citenamefont {Caldwell}\ \emph {et~al.}(2018)\citenamefont
  {Caldwell}, \citenamefont {Didier}, \citenamefont {Ryan}, \citenamefont
  {Sete}, \citenamefont {Hudson}, \citenamefont {Karalekas}, \citenamefont
  {Manenti}, \citenamefont {da~Silva}, \citenamefont {Sinclair}, \citenamefont
  {Acala} \emph {et~al.}}]{caldwell2018parametrically}%
  \BibitemOpen
  \bibfield  {author} {\bibinfo {author} {\bibfnamefont {S.}~\bibnamefont
  {Caldwell}}, \bibinfo {author} {\bibfnamefont {N.}~\bibnamefont {Didier}},
  \bibinfo {author} {\bibfnamefont {C.}~\bibnamefont {Ryan}}, \bibinfo {author}
  {\bibfnamefont {E.}~\bibnamefont {Sete}}, \bibinfo {author} {\bibfnamefont
  {A.}~\bibnamefont {Hudson}}, \bibinfo {author} {\bibfnamefont
  {P.}~\bibnamefont {Karalekas}}, \bibinfo {author} {\bibfnamefont
  {R.}~\bibnamefont {Manenti}}, \bibinfo {author} {\bibfnamefont
  {M.}~\bibnamefont {da~Silva}}, \bibinfo {author} {\bibfnamefont
  {R.}~\bibnamefont {Sinclair}}, \bibinfo {author} {\bibfnamefont
  {E.}~\bibnamefont {Acala}}, \emph {et~al.},\ }\href@noop {} {\bibfield
  {journal} {\bibinfo  {journal} {Physical Review Applied}\ }\textbf {\bibinfo
  {volume} {10}},\ \bibinfo {pages} {034050} (\bibinfo {year}
  {2018})}\BibitemShut {NoStop}%
\bibitem [{\citenamefont {Deutsch}\ \emph {et~al.}(1995)\citenamefont
  {Deutsch}, \citenamefont {Barenco},\ and\ \citenamefont
  {Ekert}}]{Deutsch1995-ax}%
  \BibitemOpen
  \bibfield  {author} {\bibinfo {author} {\bibfnamefont {D.}~\bibnamefont
  {Deutsch}}, \bibinfo {author} {\bibfnamefont {A.}~\bibnamefont {Barenco}},\
  and\ \bibinfo {author} {\bibfnamefont {A.}~\bibnamefont {Ekert}},\
  }\href@noop {} {\bibfield  {journal} {\bibinfo  {journal} {Proc., Math. phys.
  sci.}\ }\textbf {\bibinfo {volume} {449}},\ \bibinfo {pages} {669} (\bibinfo
  {year} {1995})}\BibitemShut {NoStop}%
\bibitem [{\citenamefont {Lloyd}(1995)}]{Lloyd1995-fq}%
  \BibitemOpen
  \bibfield  {author} {\bibinfo {author} {\bibfnamefont {S.}~\bibnamefont
  {Lloyd}},\ }\href@noop {} {\bibfield  {journal} {\bibinfo  {journal} {Phys.
  Rev. Lett.}\ }\textbf {\bibinfo {volume} {75}},\ \bibinfo {pages} {346}
  (\bibinfo {year} {1995})}\BibitemShut {NoStop}%
\bibitem [{\citenamefont {Bremner}\ \emph {et~al.}(2002)\citenamefont
  {Bremner}, \citenamefont {Dawson}, \citenamefont {Dodd}, \citenamefont
  {Gilchrist}, \citenamefont {Harrow}, \citenamefont {Mortimer}, \citenamefont
  {Nielsen},\ and\ \citenamefont {Osborne}}]{Bremner2002-eh}%
  \BibitemOpen
  \bibfield  {author} {\bibinfo {author} {\bibfnamefont {M.~J.}\ \bibnamefont
  {Bremner}}, \bibinfo {author} {\bibfnamefont {C.~M.}\ \bibnamefont {Dawson}},
  \bibinfo {author} {\bibfnamefont {J.~L.}\ \bibnamefont {Dodd}}, \bibinfo
  {author} {\bibfnamefont {A.}~\bibnamefont {Gilchrist}}, \bibinfo {author}
  {\bibfnamefont {A.~W.}\ \bibnamefont {Harrow}}, \bibinfo {author}
  {\bibfnamefont {D.}~\bibnamefont {Mortimer}}, \bibinfo {author}
  {\bibfnamefont {M.~A.}\ \bibnamefont {Nielsen}},\ and\ \bibinfo {author}
  {\bibfnamefont {T.~J.}\ \bibnamefont {Osborne}},\ }\href@noop {} {\bibfield
  {journal} {\bibinfo  {journal} {Phys. Rev. Lett.}\ }\textbf {\bibinfo
  {volume} {89}},\ \bibinfo {pages} {247902} (\bibinfo {year}
  {2002})}\BibitemShut {NoStop}%
\bibitem [{\citenamefont {Barenco}\ \emph {et~al.}(1995)\citenamefont
  {Barenco}, \citenamefont {Bennett}, \citenamefont {Cleve}, \citenamefont
  {DiVincenzo}, \citenamefont {Margolus}, \citenamefont {Shor}, \citenamefont
  {Sleator}, \citenamefont {Smolin},\ and\ \citenamefont
  {Weinfurter}}]{Barenco1995-ro}%
  \BibitemOpen
  \bibfield  {author} {\bibinfo {author} {\bibfnamefont {A.}~\bibnamefont
  {Barenco}}, \bibinfo {author} {\bibfnamefont {C.~H.}\ \bibnamefont
  {Bennett}}, \bibinfo {author} {\bibfnamefont {R.}~\bibnamefont {Cleve}},
  \bibinfo {author} {\bibfnamefont {D.~P.}\ \bibnamefont {DiVincenzo}},
  \bibinfo {author} {\bibfnamefont {N.}~\bibnamefont {Margolus}}, \bibinfo
  {author} {\bibfnamefont {P.}~\bibnamefont {Shor}}, \bibinfo {author}
  {\bibfnamefont {T.}~\bibnamefont {Sleator}}, \bibinfo {author} {\bibfnamefont
  {J.~A.}\ \bibnamefont {Smolin}},\ and\ \bibinfo {author} {\bibfnamefont
  {H.}~\bibnamefont {Weinfurter}},\ }\href@noop {} {\bibfield  {journal}
  {\bibinfo  {journal} {Phys. Rev. A}\ }\textbf {\bibinfo {volume} {52}},\
  \bibinfo {pages} {3457} (\bibinfo {year} {1995})}\BibitemShut {NoStop}%
\bibitem [{\citenamefont {Cleve}\ \emph {et~al.}(1998)\citenamefont {Cleve},
  \citenamefont {Ekert}, \citenamefont {Macchiavello},\ and\ \citenamefont
  {Mosca}}]{Cleve1998-mr}%
  \BibitemOpen
  \bibfield  {author} {\bibinfo {author} {\bibfnamefont {R.}~\bibnamefont
  {Cleve}}, \bibinfo {author} {\bibfnamefont {A.}~\bibnamefont {Ekert}},
  \bibinfo {author} {\bibfnamefont {C.}~\bibnamefont {Macchiavello}},\ and\
  \bibinfo {author} {\bibfnamefont {M.}~\bibnamefont {Mosca}},\ }\href@noop {}
  {\bibfield  {journal} {\bibinfo  {journal} {Proceedings of the Royal Society
  of London. Series A: Mathematical, Physical and Engineering Sciences}\
  }\textbf {\bibinfo {volume} {454}},\ \bibinfo {pages} {339} (\bibinfo {year}
  {1998})}\BibitemShut {NoStop}%
\bibitem [{\citenamefont {Khatri}\ \emph {et~al.}(2019)\citenamefont {Khatri},
  \citenamefont {LaRose}, \citenamefont {Poremba}, \citenamefont {Cincio},
  \citenamefont {Sornborger},\ and\ \citenamefont {Coles}}]{khatri2019quantum}%
  \BibitemOpen
  \bibfield  {author} {\bibinfo {author} {\bibfnamefont {S.}~\bibnamefont
  {Khatri}}, \bibinfo {author} {\bibfnamefont {R.}~\bibnamefont {LaRose}},
  \bibinfo {author} {\bibfnamefont {A.}~\bibnamefont {Poremba}}, \bibinfo
  {author} {\bibfnamefont {L.}~\bibnamefont {Cincio}}, \bibinfo {author}
  {\bibfnamefont {A.~T.}\ \bibnamefont {Sornborger}},\ and\ \bibinfo {author}
  {\bibfnamefont {P.~J.}\ \bibnamefont {Coles}},\ }\href@noop {} {\bibfield
  {journal} {\bibinfo  {journal} {Quantum}\ }\textbf {\bibinfo {volume} {3}},\
  \bibinfo {pages} {140} (\bibinfo {year} {2019})}\BibitemShut {NoStop}%
\bibitem [{\citenamefont {Amy}\ \emph {et~al.}(2013)\citenamefont {Amy},
  \citenamefont {Maslov}, \citenamefont {Mosca},\ and\ \citenamefont
  {Roetteler}}]{amy2013meet}%
  \BibitemOpen
  \bibfield  {author} {\bibinfo {author} {\bibfnamefont {M.}~\bibnamefont
  {Amy}}, \bibinfo {author} {\bibfnamefont {D.}~\bibnamefont {Maslov}},
  \bibinfo {author} {\bibfnamefont {M.}~\bibnamefont {Mosca}},\ and\ \bibinfo
  {author} {\bibfnamefont {M.}~\bibnamefont {Roetteler}},\ }\href@noop {}
  {\bibfield  {journal} {\bibinfo  {journal} {IEEE Transactions on
  Computer-Aided Design of Integrated Circuits and Systems}\ }\textbf {\bibinfo
  {volume} {32}},\ \bibinfo {pages} {818} (\bibinfo {year} {2013})}\BibitemShut
  {NoStop}%
\bibitem [{\citenamefont {Han}\ \emph {et~al.}(2020)\citenamefont {Han},
  \citenamefont {Cai}, \citenamefont {Li}, \citenamefont {Wu}, \citenamefont
  {Ma}, \citenamefont {Ma}, \citenamefont {Wang}, \citenamefont {Zhang},
  \citenamefont {Song},\ and\ \citenamefont {Duan}}]{PhysRevA.102.022619}%
  \BibitemOpen
  \bibfield  {author} {\bibinfo {author} {\bibfnamefont {X.~Y.}\ \bibnamefont
  {Han}}, \bibinfo {author} {\bibfnamefont {T.~Q.}\ \bibnamefont {Cai}},
  \bibinfo {author} {\bibfnamefont {X.~G.}\ \bibnamefont {Li}}, \bibinfo
  {author} {\bibfnamefont {Y.~K.}\ \bibnamefont {Wu}}, \bibinfo {author}
  {\bibfnamefont {Y.~W.}\ \bibnamefont {Ma}}, \bibinfo {author} {\bibfnamefont
  {Y.~L.}\ \bibnamefont {Ma}}, \bibinfo {author} {\bibfnamefont {J.~H.}\
  \bibnamefont {Wang}}, \bibinfo {author} {\bibfnamefont {H.~Y.}\ \bibnamefont
  {Zhang}}, \bibinfo {author} {\bibfnamefont {Y.~P.}\ \bibnamefont {Song}},\
  and\ \bibinfo {author} {\bibfnamefont {L.~M.}\ \bibnamefont {Duan}},\ }\href
  {https://doi.org/10.1103/PhysRevA.102.022619} {\bibfield  {journal} {\bibinfo
   {journal} {Phys. Rev. A}\ }\textbf {\bibinfo {volume} {102}},\ \bibinfo
  {pages} {022619} (\bibinfo {year} {2020})}\BibitemShut {NoStop}%
\bibitem [{\citenamefont {Ghosh}\ \emph {et~al.}(2013)\citenamefont {Ghosh},
  \citenamefont {Galiautdinov}, \citenamefont {Zhou}, \citenamefont {Korotkov},
  \citenamefont {Martinis},\ and\ \citenamefont {Geller}}]{PhysRevA.87.022309}%
  \BibitemOpen
  \bibfield  {author} {\bibinfo {author} {\bibfnamefont {J.}~\bibnamefont
  {Ghosh}}, \bibinfo {author} {\bibfnamefont {A.}~\bibnamefont {Galiautdinov}},
  \bibinfo {author} {\bibfnamefont {Z.}~\bibnamefont {Zhou}}, \bibinfo {author}
  {\bibfnamefont {A.~N.}\ \bibnamefont {Korotkov}}, \bibinfo {author}
  {\bibfnamefont {J.~M.}\ \bibnamefont {Martinis}},\ and\ \bibinfo {author}
  {\bibfnamefont {M.~R.}\ \bibnamefont {Geller}},\ }\href
  {https://doi.org/10.1103/PhysRevA.87.022309} {\bibfield  {journal} {\bibinfo
  {journal} {Phys. Rev. A}\ }\textbf {\bibinfo {volume} {87}},\ \bibinfo
  {pages} {022309} (\bibinfo {year} {2013})}\BibitemShut {NoStop}%
\bibitem [{\citenamefont {Venturelli}\ \emph {et~al.}(2018)\citenamefont
  {Venturelli}, \citenamefont {Do}, \citenamefont {Rieffel},\ and\
  \citenamefont {Frank}}]{venturelli2018compiling}%
  \BibitemOpen
  \bibfield  {author} {\bibinfo {author} {\bibfnamefont {D.}~\bibnamefont
  {Venturelli}}, \bibinfo {author} {\bibfnamefont {M.}~\bibnamefont {Do}},
  \bibinfo {author} {\bibfnamefont {E.}~\bibnamefont {Rieffel}},\ and\ \bibinfo
  {author} {\bibfnamefont {J.}~\bibnamefont {Frank}},\ }\href@noop {}
  {\bibfield  {journal} {\bibinfo  {journal} {Quantum Science and Technology}\
  }\textbf {\bibinfo {volume} {3}},\ \bibinfo {pages} {025004} (\bibinfo {year}
  {2018})}\BibitemShut {NoStop}%
\bibitem [{\citenamefont {Cincio}\ \emph {et~al.}(2018)\citenamefont {Cincio},
  \citenamefont {Suba{\c{s}}{\i}}, \citenamefont {Sornborger},\ and\
  \citenamefont {Coles}}]{cincio2018learning}%
  \BibitemOpen
  \bibfield  {author} {\bibinfo {author} {\bibfnamefont {L.}~\bibnamefont
  {Cincio}}, \bibinfo {author} {\bibfnamefont {Y.}~\bibnamefont
  {Suba{\c{s}}{\i}}}, \bibinfo {author} {\bibfnamefont {A.~T.}\ \bibnamefont
  {Sornborger}},\ and\ \bibinfo {author} {\bibfnamefont {P.~J.}\ \bibnamefont
  {Coles}},\ }\href@noop {} {\bibfield  {journal} {\bibinfo  {journal} {New
  Journal of Physics}\ }\textbf {\bibinfo {volume} {20}},\ \bibinfo {pages}
  {113022} (\bibinfo {year} {2018})}\BibitemShut {NoStop}%
\bibitem [{\citenamefont {Zahedinejad}\ \emph {et~al.}(2016)\citenamefont
  {Zahedinejad}, \citenamefont {Ghosh},\ and\ \citenamefont
  {Sanders}}]{zahedinejad2016designing}%
  \BibitemOpen
  \bibfield  {author} {\bibinfo {author} {\bibfnamefont {E.}~\bibnamefont
  {Zahedinejad}}, \bibinfo {author} {\bibfnamefont {J.}~\bibnamefont {Ghosh}},\
  and\ \bibinfo {author} {\bibfnamefont {B.~C.}\ \bibnamefont {Sanders}},\
  }\href@noop {} {\bibfield  {journal} {\bibinfo  {journal} {Physical Review
  Applied}\ }\textbf {\bibinfo {volume} {6}},\ \bibinfo {pages} {054005}
  (\bibinfo {year} {2016})}\BibitemShut {NoStop}%
\bibitem [{\citenamefont {Booth}\ \emph {et~al.}(2018)\citenamefont {Booth},
  \citenamefont {Do}, \citenamefont {Beck}, \citenamefont {Rieffel},
  \citenamefont {Venturelli},\ and\ \citenamefont
  {Frank}}]{booth2018comparing}%
  \BibitemOpen
  \bibfield  {author} {\bibinfo {author} {\bibfnamefont {K.~E.}\ \bibnamefont
  {Booth}}, \bibinfo {author} {\bibfnamefont {M.}~\bibnamefont {Do}}, \bibinfo
  {author} {\bibfnamefont {J.~C.}\ \bibnamefont {Beck}}, \bibinfo {author}
  {\bibfnamefont {E.}~\bibnamefont {Rieffel}}, \bibinfo {author} {\bibfnamefont
  {D.}~\bibnamefont {Venturelli}},\ and\ \bibinfo {author} {\bibfnamefont
  {J.}~\bibnamefont {Frank}},\ }in\ \href@noop {} {\emph {\bibinfo {booktitle}
  {Twenty-Eighth international conference on automated planning and
  scheduling}}}\ (\bibinfo {year} {2018})\BibitemShut {NoStop}%
\bibitem [{\citenamefont {Maslov}(2016)}]{maslov2016advantages}%
  \BibitemOpen
  \bibfield  {author} {\bibinfo {author} {\bibfnamefont {D.}~\bibnamefont
  {Maslov}},\ }\href@noop {} {\bibfield  {journal} {\bibinfo  {journal}
  {Physical Review A}\ }\textbf {\bibinfo {volume} {93}},\ \bibinfo {pages}
  {022311} (\bibinfo {year} {2016})}\BibitemShut {NoStop}%
\bibitem [{\citenamefont {Martinez}\ \emph {et~al.}(2016)\citenamefont
  {Martinez}, \citenamefont {Monz}, \citenamefont {Nigg}, \citenamefont
  {Schindler},\ and\ \citenamefont {Blatt}}]{martinez2016compiling}%
  \BibitemOpen
  \bibfield  {author} {\bibinfo {author} {\bibfnamefont {E.~A.}\ \bibnamefont
  {Martinez}}, \bibinfo {author} {\bibfnamefont {T.}~\bibnamefont {Monz}},
  \bibinfo {author} {\bibfnamefont {D.}~\bibnamefont {Nigg}}, \bibinfo {author}
  {\bibfnamefont {P.}~\bibnamefont {Schindler}},\ and\ \bibinfo {author}
  {\bibfnamefont {R.}~\bibnamefont {Blatt}},\ }\href@noop {} {\bibfield
  {journal} {\bibinfo  {journal} {New Journal of Physics}\ }\textbf {\bibinfo
  {volume} {18}},\ \bibinfo {pages} {063029} (\bibinfo {year}
  {2016})}\BibitemShut {NoStop}%
\bibitem [{\citenamefont {Younis}\ \emph {et~al.}(2020)\citenamefont {Younis},
  \citenamefont {Sen}, \citenamefont {Yelick},\ and\ \citenamefont
  {Iancu}}]{younis2020qfast}%
  \BibitemOpen
  \bibfield  {author} {\bibinfo {author} {\bibfnamefont {E.}~\bibnamefont
  {Younis}}, \bibinfo {author} {\bibfnamefont {K.}~\bibnamefont {Sen}},
  \bibinfo {author} {\bibfnamefont {K.}~\bibnamefont {Yelick}},\ and\ \bibinfo
  {author} {\bibfnamefont {C.}~\bibnamefont {Iancu}},\ }\href@noop {}
  {\bibfield  {journal} {\bibinfo  {journal} {arXiv preprint arXiv:2003.04462}\
  } (\bibinfo {year} {2020})}\BibitemShut {NoStop}%
\bibitem [{\citenamefont {Nakanishi}\ \emph {et~al.}(2020)\citenamefont
  {Nakanishi}, \citenamefont {Fujii},\ and\ \citenamefont
  {Todo}}]{Nakanishi2020}%
  \BibitemOpen
  \bibfield  {author} {\bibinfo {author} {\bibfnamefont {K.~M.}\ \bibnamefont
  {Nakanishi}}, \bibinfo {author} {\bibfnamefont {K.}~\bibnamefont {Fujii}},\
  and\ \bibinfo {author} {\bibfnamefont {S.}~\bibnamefont {Todo}},\ }\href@noop
  {} {\bibfield  {journal} {\bibinfo  {journal} {Phys. Rev. Research}\ }\textbf
  {\bibinfo {volume} {2}},\ \bibinfo {pages} {043158} (\bibinfo {year}
  {2020})}\BibitemShut {NoStop}%
\bibitem [{\citenamefont {Schuch}\ and\ \citenamefont
  {Siewert}(2003)}]{Schuch2003-ze}%
  \BibitemOpen
  \bibfield  {author} {\bibinfo {author} {\bibfnamefont {N.}~\bibnamefont
  {Schuch}}\ and\ \bibinfo {author} {\bibfnamefont {J.}~\bibnamefont
  {Siewert}},\ }\href@noop {} {\bibfield  {journal} {\bibinfo  {journal} {Phys.
  Rev. Lett.}\ }\textbf {\bibinfo {volume} {91}},\ \bibinfo {pages} {027902}
  (\bibinfo {year} {2003})}\BibitemShut {NoStop}%
\bibitem [{\citenamefont {Schuch}\ and\ \citenamefont
  {Siewert}(2002)}]{Schuch_undated-yw}%
  \BibitemOpen
  \bibfield  {author} {\bibinfo {author} {\bibfnamefont {N.}~\bibnamefont
  {Schuch}}\ and\ \bibinfo {author} {\bibfnamefont {J.}~\bibnamefont
  {Siewert}},\ }\emph {\bibinfo {title} {{Implementation of quantum algorithms
  with Josephson charge qubits}}},\ \href@noop {} {Ph.D. thesis},\ \bibinfo
  {school} {Universit\"{a}t Regensburg} (\bibinfo {year} {2002})\BibitemShut
  {NoStop}%
\bibitem [{\citenamefont {Gwinner}\ \emph {et~al.}(2020)\citenamefont
  {Gwinner}, \citenamefont {Bria{\'n}ski}, \citenamefont {Burkot},
  \citenamefont {Czerwi{\'n}ski},\ and\ \citenamefont
  {Hlembotskyi}}]{gwinner2020benchmarking}%
  \BibitemOpen
  \bibfield  {author} {\bibinfo {author} {\bibfnamefont {J.}~\bibnamefont
  {Gwinner}}, \bibinfo {author} {\bibfnamefont {M.}~\bibnamefont
  {Bria{\'n}ski}}, \bibinfo {author} {\bibfnamefont {W.}~\bibnamefont
  {Burkot}}, \bibinfo {author} {\bibfnamefont {{\L}.}~\bibnamefont
  {Czerwi{\'n}ski}},\ and\ \bibinfo {author} {\bibfnamefont {V.}~\bibnamefont
  {Hlembotskyi}},\ }\href@noop {} {\bibfield  {journal} {\bibinfo  {journal}
  {arXiv preprint arXiv:2007.06539}\ } (\bibinfo {year} {2020})}\BibitemShut
  {NoStop}%
\end{thebibliography}%

\onecolumngrid
\clearpage
\appendix

\section{Derivation of Eq.~(\ref{eq:trU})}\label{app:trU}

\begin{align}
    \min_\phi \norm{e^{i\phi}V_T^\dag VP - P}_F^2
    &= \min_\phi \qty(\tr[\qty(\qty(e^{i\phi}V_T^\dag VP)^\dag - P)\qty\Big(e^{i\phi}V_T^\dag VP - P)]) \notag\\
    &= \min_\phi \qty(2D - 2\Re(\tr[e^{i\phi}V_T^\dag VP]))\quad\qty(\because \tr[P^2]=\tr[P]=D) \notag\\
    &= 2D - 2\max_\phi \qty(\Re(\tr[e^{i\phi}V_T^\dag VP])) \notag\\
    &= 2D - 2\max_\phi \qty(\Re(e^{i\phi} \tr[V_T^\dag VP])) \notag\\
    &= 2D - 2\qty|\tr[V_T^\dag VP]|
\end{align}

\section{Derivation of Eq.~(\ref{eq:sinU})}\label{app:sinU}

\begin{align}
    \tr[V_T^\dag \tilde{V}_n(\theta)P]
    &= \tr[V_T^\dag \tilde{V}_n(0)P \cos\frac{\theta}{2} + V_T^\dag \tilde{V}_n(\pi)P \sin\frac{\theta}{2}] \notag\\
    &= \tr[V_T^\dag \tilde{V}_n(0)P] \cos\frac{\theta}{2} + \tr[V_T^\dag \tilde{V}_n(\pi)P] \sin\frac{\theta}{2} \notag\\
    &= t_0 \cos\frac{\theta}{2} + t_\pi \sin\frac{\theta}{2},
\end{align}
where $t_0:=\tr[V_T^\dag \tilde{V}_n(0)P]$, $t_\pi:=\tr[V_T^\dag \tilde{V}_n(\pi)P]$.
Then,
\begin{align}
    \qty|\tr[V_T^\dag \tilde{V}_n(\theta)P]|^2
    &= \qty|t_0 \cos\frac{\theta}{2} + t_\pi \sin\frac{\theta}{2}|^2 \notag\\
    &= \qty|t_0|^2 \cos^2\frac{\theta}{2} + \qty|t_\pi|^2 \sin^2\frac{\theta}{2}
    + \qty\big(t_0 t_\pi^* + t_0^* t_\pi)\cos\frac{\theta}{2}\sin\frac{\theta}{2} \notag\\
    &= \frac{\qty|t_0|^2 - \qty|t_\pi|^2}{2} \cos\theta
    + \frac{t_0 t_\pi^* + t_0^* t_\pi}{2} \sin\theta
    + \frac{\qty|t_0|^2 + \qty|t_\pi|^2}{2}
\end{align}

\end{document}